\documentstyle{l-aa}

\begin{document}
\title{Binary stars in globular clusters: detection of a binary sequence 
in NGC 2808?$^{\dag}$}
\author{F.R. Ferraro \inst{1}, E. Carretta \inst{1}, F. Fusi Pecci \inst{1,2}, 
A. Zamboni \inst{1}}
\institute{(1) Osservatorio Astronomico di Bologna, Via 
Zamboni 33, 40126 Bologna, Italy\\
(2) Stazione Astronomica di Cagliari, I-09012 Capoterra, Italy}
\thesaurus{08.02.3, 08.08.1, 08.16.3, 10.07.3 } 
\offprints{ferraro@astbo3.bo.astro.it}
\date{$^{\dag}$Based on observations obtained at 
the European Southern Observatory, La Silla, Chile}
\maketitle
\markboth{F.R. Ferraro et al.}{Binary stars in globular clusters.}
\begin{abstract} 
A CCD $V$,$V-I$ colour-magnitude diagram (CMD) of the Galactic globular 
cluster NGC 2808 has been obtained with the ESO-NTT, reaching down to 
$V \sim 24$.

The highly populated Main Sequence (MS) presents a significant broadening 
redward to the MS ridge line, larger than expected from photometric errors
alone, which could be interpreted as due to binary candidates.

\keywords{Cluster: globular: NGC 2808, 
Techniques: photometry,
stars: Population II, binaries: general}
 
\end{abstract}

\section{Introduction}
There are various (direct and indirect, but always difficult) approaches 
followed so far to detect binaries in Galactic Globular Custers (GGCs)
(see Hut et al. 1992 and Bailyn 1995 for a discussion). 
Most of them have been unsuccesful, but 
this may well be due to the fact that the surviving binaries are likely 
segregated in the inner cluster regions, hardly observable from the ground 
because of crowding.

One possible path to explore consists in the study of the intrinsic
width of the Main Sequence (MS), looking for some sort of ``parallel"
sequence due to the photometric combination of the binary components (see
Romani and Weimberg 1991). This approach requires the achievement of
a very accurate photometry (to $\sim 0.01 - 0.02$ mag) for a very wide
sample of faint MS stars. On the other hand, other methods imply
time-consuming and distance-constrained spectroscopic observations, usually
sensitive only to the brighter objects with present day instrumentation.

The true fraction and the physical nature of binary stars in a GGC is a
clue element that brings information on a variety of topics like the 
environment in which the cluster was born, the details of the star
formation, and the subsequent internal dynamical evolution. 
While theoretical models are presently only poorly
constrained by any observational (and quantitative) data-set, it is now 
well established that a strict connection does  exist between stellar
and dynamical evolution of a GC star, as revealed for instance by the 
morphology of the evolved branches, in particular the Horizonthal Branch 
(HB; see e.g. Fusi Pecci et al. 1993 for discussion and references).

NGC 2808 ($\alpha_{1950} = 9^h 10.9^m, \delta_{1950} = -64^o 39'$) is 
one of the most interesting objects to study, in this respect, since its
very high concentration is likely to enhance any effect due to
the dynamical evolution of binary or multiple star systems.
This cluster is also the typical template of a growing class 
of GGCs with a net, bimodal HB morphology (see Harris 1975, Ferraro et al.
1990-F90, and Rood et al. 1993). This, in turn, could well be heavily 
affected by the evolution of binary systems that at later stages
can populate both the red and the blue HB extremes (Fusi Pecci 
et al. 1993). Further, to explain its bimodal HB,  van den Bergh (1997) 
has recently suggested that this cluster might have formed by mergers (but
see Catelan 1997). Any new (though preliminar) information can thus
be useful to have a deeper insight on its properties.

In this research note we present a very preliminary evidence of a secondary
sequence running parallel to the MS of this cluster that, if confirmed,
could be interpreted as composed of unresolved, candidate binary systems.

\section{Observations and data reduction}

A set of deep $V$ and $I$ frames (3 exposures of 600 secs in each filter) were
acquired under good seeing conditions (0.7-0.9" FWHM) on January 1995 
at the NTT telescope (ESO, La Silla, Chile) with the EMMI camera equipped 
with a 2048 $\times$ 2048 CCD detector. The image scale is 0.25 arcsec/pixel, 
yielding a total field of view of 8.5' $\times$ 8.5'.
In order to avoid severe crowding conditions, the results reported here
are based only on a sub-field of about 4' $\times$ 4', whose center lies about
6' southward and 2' eastward away from the cluster center (see Figure 1). 

All reductions were performed using the package ROMAFOT
(Buonanno et al. 1983). Details of the image analysis are very similar
to those described elsewhere (see F90, Ferraro et al. 1992) and will not
repeated here. The data-files are available upon request.

The instrumental magnitudes (from profile-fitting) were first tied to fixed
aperture photometry and then referred to the photometric Johnson
system using 15 stars in the Landolt (1992) standard fields Rubin 149 and SA98.
The resulting calibration equations are: 
$$ V = v + 0.024 (v-i) + 23.793 $$ 
$$ I = i -0.077 (v-i) + 23.260$$
where $V$, $I$ are standard magnitudes and $v$, $i$ indicate the
instrumental ones. Atmospheric extinction was taken into account 
by means of average extinction curves appropriate for La Silla.

Based on comparison of 70 stars in common, we note here that our new 
calibration is fainter (0.12 mag) in $V$ than previously obtained by F90; 
no comparison was feasible in the $I$ band. 

\begin{figure}[tbhp]
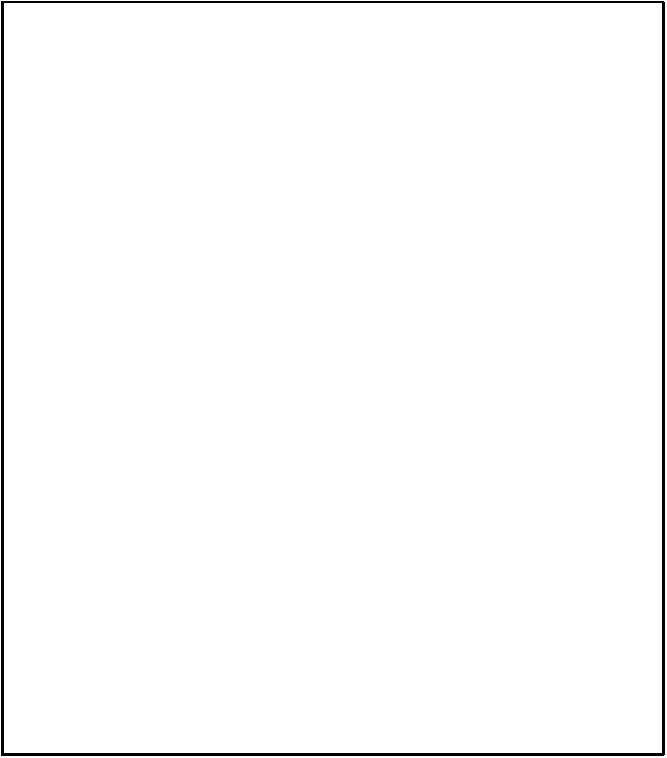
 
\picplace{10.0 cm} 
\caption[]{The map of NGC 2808 with the location of the sub-frame of
about 4' $\times$ 4' analyzed in this note.}
\label{f:fig1}
\end{figure}

\section{Results}
Figure 2 shows the ($V$, $V-I$) CMD for 3174 stars detected in the selected
sub-frame. Both the turn-off (TO) region
and MS are well defined. However, {\it the most intriguing evidence is
the presence of a clear broadening above and to the red of the MS, 
which can be interpreted as a secondary sequence, running parallel to 
the MS, but at brighter magnitudes and redder colour.}

In Figure 2, to make more evident this feature, we marked the claimed 
parallel sequence with a dashed line.
The distance in magnitude of this ``secondary" sequence from the MS
is somewhat less than 0.75 mag. 
Note that equal component binaries would lie at the same colour and 0.75 mag
brighter than the two individual members (Romani and Weinberg 1991), and
this magnitude difference should thus be considered as an upper limit.
Consequently, taking also into account the photometric errors, the 
average difference here measured for the two parallel sequences is
substantially compatible with the above limits.

\begin{figure}[tbhp]
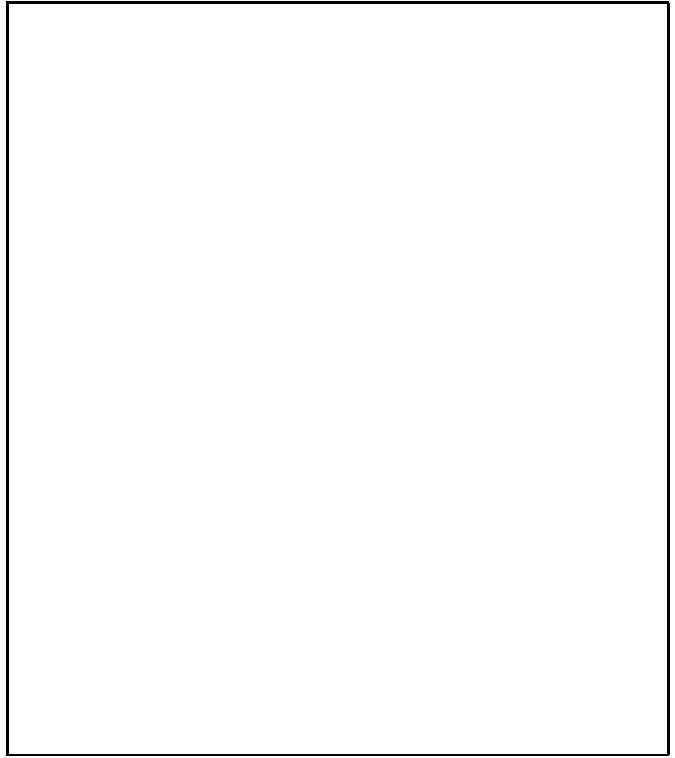
 
\picplace{10.0 cm} 
\caption[]{$V$,$V-I$ CMD of NGC 2808; errorbars for the
photometry are plotted on the left. Mean ridge lines for the MS
(solid line) and the candidate binary sequence (dashed line)
are overimposed to the data.}
\label{f:fig2}
\end{figure} 

To test the existence of the candidate binary sequence we 
have considered various luminosity bins along the MS 
(between 21 and 22.5, see Table 1). 
Then for all the stars in each bin we derived the distribution of the
distances ($\delta x$) from the MS ridge line (MSRL). The observable 
($\delta x$) is defined as the geometrical distance in the ($V$, $V-I$)-plane
of each star from the adopted MSRL, with $\delta x > 0$ or $\delta x < 0$
if the star is redder or bluer than the MSRL, respectively.

In Figure 3, the distributions of the ($\delta x$)-values in the 
ranges $V = 21.5 \div 22.0$ and $V = 21.0 \div 22.0$ are shown for 
illustration and clarity.

\begin{table}
\begin{center}
\caption{ K-S test at different Magnitude ranges considered along the MS}
\begin{tabular}{lcc}
\hline\hline
\\
Section & Mag. range & K-S \\
\\
\hline
\\
A       & 21.0$-$22.0& 0.7\% \\
B       & 21.5$-$22.0& 0.5\% \\
C       & 21.5$-$22.5& 3.0\% \\
\\
\hline
\end{tabular}
\end{center}
\end{table}

\begin{figure}[tbhp]
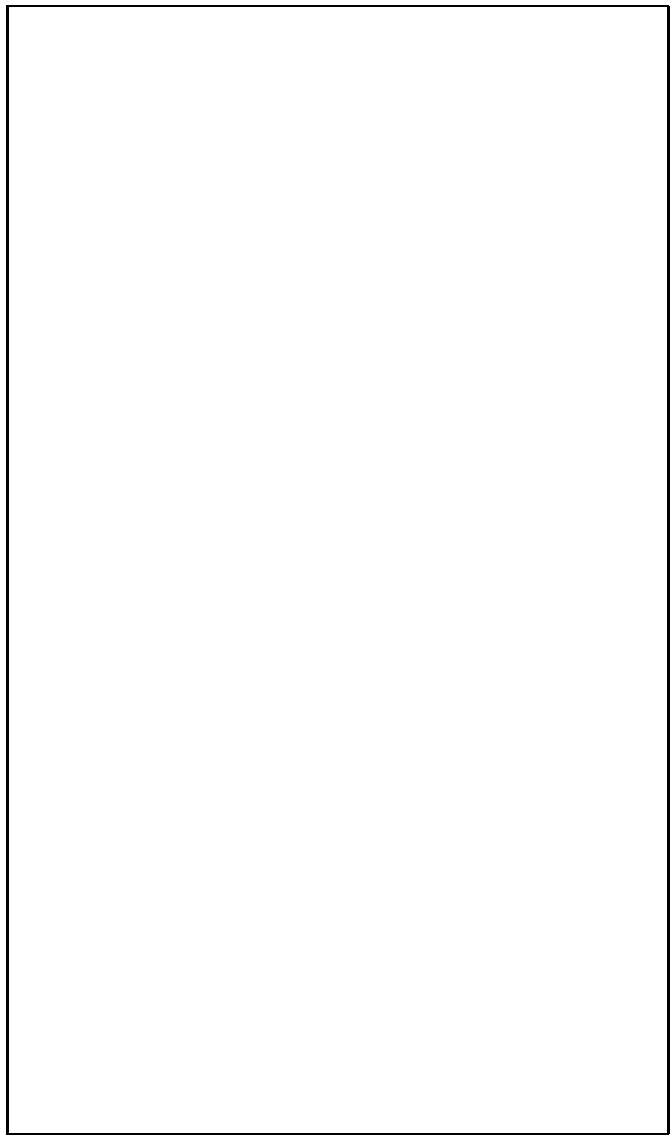
 
\picplace{15.0 cm} 
\caption[]{Histograms of the distances of stars from the MS
mean ridge line in NGC 2808 (dots); negative values refer to stars 
bluer than the ridge line. Gaussian fits are overimposed to the data.
Panel (a): bin from $V = 21.5$ to $V = 22.0$, panel (b): bin
from $V = 21.0$ to $V = 22.0$.}
\label{f:fig3}
\end{figure} 

As can easily be seen, both panels in Figure 3 show that the distributions 
of the ($\delta x$)-values are clearly asymmetric.
In fact, besides the main peak
centered on the MS, there is a quite evident secondary peak, displaced by
about $\delta$x = 0.12 from the main one. In our view, this is a 
direct evidence for the presence of a secondary sequence along the MS as 
the asymmetric shape itself of these distributions is a proof that the
broadening of the MS is not just due to photometric errors.

Before discussing at any level the possible origin of the stars located
in the parallel sequence, we have tested the statistical significance
of this claimed feature. To this aim we have compared the number of stars
distributed over equal intervals in distance ($|\delta x|$) from the MSRL
lying in the blue and red side of the distributions.

Over the magnitude interval $21.5 < V < 22.0$ (where the effect is best
evident), the number of stars within $|\delta x| < 0.1$ from the MSRL is
almost the same (within 1 $\sigma$) on both side of the distribution
(138 stars on the blue side and 159 on the red one, respectively). On the
contrary, if one consider the range $0.1 < |\delta x| < 0.2$, one finds only
15 stars bluer than the MSRL to be compared with 48 objects located on the 
red side. Hence, the number of red stars is more than 3 times that found 
in the blue side of the distribution.

Moreover, a Kolmogorov-Smirnov (KS) test applied to evaluate the
significance of the detected difference in the red and blue distributions 
yields a very low probability that the two distributions have been extracted 
from the same $parent$ distribution. In fact, as can be seen in the last
column of Table 1 for the 3 intervals of magnitudes between $V = 21.0$ and
$V = 22.5$, the probability that the presence of the secondary peak is
not due to a statistical fluctuation ranges from $\sim 97\%$ (2.2 $\sigma$) to
$\sim 99.5\%$ (2.8 $\sigma$). 

The ($V$,$V-I$)-CMD shown in Figure 2 and the statistical tests carried
out above indicate thus that several stars in the MS of NGC 2808 lie on a 
secondary sequence running parallel and redward to the MS.
From the areas of the interpolating Gaussian fits (see Figure 3)
we can also obtain a (rough) estimate of the percentage of stars populating
the two sequences over the total number of MS stars.
In the different magnitude bins considered above we obtain for the
parallel sequences figures ranging from 20.0 to 28.0 $\%$, with a mean 
values of $\sim (24 \pm 4) \%$.

\section{Optical and/or physical binaries?}

Concerning the possible nature of the stars populating the parallel
sequence, the label {\it unresolved binary} has to be taken with 
particular caution as it may mean either unresolved $optical$ binaries 
or unresolved $physical$ binaries. In fact, since we are making photometric 
measurements in a very crowded field, it has to be seen as quite normal 
the detection of some blended images.

On the other hand, the fraction of candidate binaries we have found 
($\sim 24\%$) is somewhat too high compared to the number of unresolved 
blends one can predict to find (just because of crowding) "making artificial
star" experiments (which would yield here values close to $\sim 10\%$).
As a consequence, one could also conclude as a working hypothesis
that a non neglegible fraction of the stars located on the 
parallel MS could represent reliable intrinsic binary candidates.

Obviously, since stellar crowding mimics the effect of the physical 
dynamical relaxation acting in the cluster and concentrates the optical blends
toward the cluster center, as does mass segregation with true (and more
massive) binary systems, it is too early to extend deeply the discussion of
the possible implications of this observational result, if confirmed.
However, it may be interesting to recall that NGC 2808 presents one of 
the most intriguing HB-morphologies found in a Galactic globular cluster 
so far. As reported in the Introduction, since the early observations
carried out by Harris (1974, 1975), it was evident that the HB of NGC 2808
is highly bimodal. Moreover, the latest HST data presented by Djorgovski 
et al. (1996) have shown that, besides the already detected bimodality, the
blue HB of NGC 2808 is actually made by 3 sub-portions, separated by quite
evident gaps.

Within this framework still so 
uncertain and worth of further exploration, the detection of a fraction
of $physical$ binaries could open new perspectives to the possible
interpretation of the detected peculiarities in the HB morphology. 
For instance, one could imagine that at least a group of the HB stars could
represent the descendants of the binary systems. The discussion of how
this could occur is however beyond the purposes of the present work, and 
we refer to the review papers by Hut et al. (1992) and Bailyn (1995) for
further analysis of this aspect.

\begin{acknowledgements}
This work has been supported by the {\it Agenzia Spaziale Italiana} (ASI) and
by the {\it Ministero della Universit\`a e della Ricerca Scientifica e 
Tecnologica} (MURST).
\end{acknowledgements}

\end{document}